\documentclass[preprint,12pt]{aastex}

\usepackage{emulateapj5}

\slugcomment{Accepted for publication in AJ}

\shorttitle{New Southern T Dwarfs}
\shortauthors{Burgasser et al.}

\begin{document}

\title{The 2MASS Wide-Field T Dwarf Search. II. Discovery of Three T Dwarfs in the Southern Hemisphere}

\author{
Adam J.\ Burgasser\altaffilmark{1,2},
Michael W.\ McElwain\altaffilmark{1},
\& J.\ Davy Kirkpatrick\altaffilmark{3}}

\altaffiltext{1}{Department of Astronomy \& Astrophysics,
University of California
at Los Angeles, Los Angeles,
CA, 90095-1562; adam@astro.ucla.edu, mcelwain@astro.ucla.edu}
\altaffiltext{2}{Hubble Fellow}
\altaffiltext{3}{Infrared Processing and Analysis Center, M/S 100-22,
California Institute of Technology, Pasadena, CA 91125; davy@ipac.caltech.edu}

\begin{abstract}
We present the discovery of three new Southern Hemisphere
T dwarfs identified in the Two Micron
All Sky Survey.  These objects, 2MASS 0348$-$6022, 2MASS 0516$-$0445, and
2MASS 2228$-$4310, have classifications T7, T5.5, and T6.5, respectively.
Using linear absolute magnitude/spectral type relations
derived from T dwarfs with measured parallaxes, we estimate spectrophotometric
distances for these discoveries; the closest, 2MASS 0348$-$6022, is likely
within 10 pc of the Sun.
Proper motions and estimated tangential velocities
are consistent with membership in the Galactic disk population.
We also list Southern Hemisphere T dwarf
candidates that were either not
found in subsequent near-infrared imaging observations and are most likely
uncatalogued minor planets, or have near-infrared spectra consistent with background
stars.
\end{abstract}

\keywords{Galaxy: solar neighborhood --- infrared: stars ---
stars: individual (2MASS J03480772$-$6022270,
2MASS J05160945$-$0445499,
2MASS J22282889$-$4310262) --- stars: low mass,
brown dwarfs}

\section{Introduction}

T dwarfs are substellar objects whose near-infrared
spectra exhibit characteristic signatures
of H$_2$O and CH$_4$ \citep{me02a,geb02}.  They comprise the coolest class of
brown dwarfs currently known, with effective temperatures
T$_{eff}$ $\lesssim$ 1300--1500 K \citep[and references therein]{me02a,dah02}.
Their atmospheric properties are therefore quite similar to
class III and IV extra-solar giant planets (EGPS; Sudarsky, Burrows, \& Pinto 2000),
but are more easily studied
without the obscuration of a bright host star.  Indeed,
EGP atmospheric models have advanced in parallel with isolated brown dwarf models,
as the latter differ only in the absence of an external radiating source
\citep{sea98,sea00,sud00,bar03}.  The observed properties of T dwarfs
have served to constrain these models.
The currently known collection of nearby
T dwarfs are also useful for studies of the substellar population
in the Solar Neighborhood and the
substellar initial mass function \citep{me01}.
Finally, as these objects do not significantly process their initial
Hydrogen supply to heavier metals\footnote{Exhaustive Deuterium and Lithium burning occur
in Solar metallicity brown dwarfs more massive than 0.013 and 0.065 M$_{\sun}$, respectively
\citep{bur01}.}, they may be used as a tracer population of the chemical
history of the Galaxy, as long as temperature, gravity, and metallicity diagnostics
can be disentangled.

We have initiated a wide-field search
for T dwarfs in the Two Micron All Sky Survey
\citep[hereafter 2MASS]{skr97}, as described in \citet[hereafter Paper I]{me03a}.
2MASS, which covers the entire sky,
is an ideal survey for finding T dwarfs in the Solar Neighborhood, as the
spectral energy distributions of these objects peak in the J, H, and K$_s$
photometric bands sampled by the survey.
The optical/near-infrared colors of T dwarfs
are also extremely red ($R-J \gtrsim 9$; Golimowski et al.\ 1998); hence,
photographic plate surveys, which also cover the whole sky,
cannot detect T dwarfs much farther than a few parsecs from the Sun
\citep{sch03}.  2MASS is particularly useful for finding T dwarfs
in the Southern Hemisphere, as it is more sensitive than the Deep Near-Infrared
Survey of the Southern Sky \citep{epc97}, and there is as yet no equivalent to
the Sloan Digital Sky Survey \citep[hereafter SDSS]{yor00} at southern latitudes.

In this article, we present the discovery of three new T dwarfs in the Southern
Hemisphere, 2MASS 0348$-$6022,
2MASS 0516$-$0445, and
2MASS 2228$-$4310\footnote{Throughout the text, we adopt shorthand notation
for 2MASS and SDSS sources, 2MASS/SDSS hhmm$\pm$ddmm.
Full designations are listed in Tables 1--3.}.
In $\S$ 2 we review the identification of these T dwarfs
through confirmation imaging
and spectroscopic observations obtained using the Ohio State
InfraRed Imager/Spectrometer \citep[hereafter OSIRIS]{dep93},
mounted on the CTIO 4m Blanco Telescope.
In $\S$ 3, we analyze the spectra of these and three other
previously identified T dwarfs, 2MASS 0243$-$2453, 2MASS 0415$-$0935,
and SDSS 0423$-$0414 \citep{me02a,geb02}, also observed with OSIRIS.
Including the latter sources into a suite of spectral standards, we
classify the new T dwarfs and estimate their spectrophotometric
distances.  We also examine the kinematics of the three discoveries.
Results are summarized in $\S$ 4.

\section{Observations of T Dwarf Candidates}

\subsection{Imaging Observations}

Selection of T dwarf candidates from the 2MASS
Working Point Source Database (WPSD) are discussed in
detail in Paper I.
Our color constraints -- $J-H \leq 0.3$ or $H-K_s \leq 0$, and no optical counterpart in
the USNO A2.0 catalogue \citep{mon98} or visible on Digitized Sky Survey (DSS)
plates -- result in some contamination by asteroids, as
discussed in \citet{me02a}.  We specifically exclude objects associated with known minor
planets in the 2MASS catalog, but uncatalogued ones are likely to be present.
To eliminate these sources, we imaged a subset of
our candidate pool with OSIRIS in the J-band during 20--24 September 2002 (UT).  Conditions during
this period ranged from clear and dry (20 and 24 September) to light cirrus and clouds
(21--23 September),
and seeing was poor ($\gtrsim 1\farcs5$)
on 20 and 23 September but excellent ($\lesssim 1\arcsec$) on the remaining nights.
Pairs of 15 sec integrations were obtained for each object,
dithered 10--20$\arcsec$ between exposures.  The sensitivity of these images
was verified to be greater than that of the 2MASS survey, which has a J-band S/N = 10
completeness limit
of 15.8 \citep{cut03}, implying that any stationary sources should have been
recovered.

As in \citet{me02a},
a number of sources were not seen in the OSIRIS images; these are listed in Table 1,
with updated designations and photometry from the
2MASS All Sky Data Release \citep[hereafter, 2MASS ADR]{cut03}.
A few sources were identified as unflagged image artifacts arising from
electronic feedback in the quadrant readout of the 2MASS NICMOS3 detectors
(``star echo''; see sect.\ IV.7 in Cutri et al.\ 2003).
Eleven sources were subsequently identified as known minor planets by
searching a $30{\arcsec}{\times}30{\arcsec}$ region around the positions
of each unconfirmed source at the corresponding epoch of observation in the
Small Body catalog maintained by the Jet Propulsion Laboratory Solar System
Dynamics Group\footnote{Asteroid identifications were made using the
Small-Body Search Tool,
\url{http://ssd.jpl.nasa.gov/cgi-bin/sb{\_}search}.}.
We note that about half of these 2MASS asteroid detections precede their discovery.
It is logical, therefore, that most of the remaining unconfirmed sources
are uncatalogued minor planets,
given that their near-infrared colors closely match those of known asteroids \citep{syk02}
and that they are generally found at low ecliptic latitudes.
Curiously, a handful of these probable minor planets
are found at relatively high ecliptic latitudes,
$25{\degr} \lesssim {\mid}{\beta}{\mid} \lesssim 40{\degr}$.
We cannot rule out the possibility of any of these
sources being eruptive or otherwise variable events.

\subsection{Spectroscopic Observations}

For sources that reappeared in the OSIRIS images,
we obtained spectra during the same run using the
OSIRIS cross-dispersed gratings.
Use of the single diffraction grating blazed at 6.6 $\micron$ with the
120 lines mm$^{-1}$ grating and f/2.8 camera provides
simultaneous, moderate resolution
($\lambda$/$\Delta\lambda$ $\sim$ 1200) spectroscopy from
1.2--2.35 $\micron$ in four orders: J (5$^{rd}$ and 6$^{th}$),
H (4$^{th}$), and K (3$^{th}$).  Resolution on the 0$\farcs$403 pixel$^{-1}$
chip ranges from 4.4 to 8.8 {\AA} pixel$^{-1}$.

Table 2 summarizes the observations, which also included
three known T dwarfs: 2MASS 0243$-$2453, 2MASS 0415$-$0935,
and SDSS 0423$-$0414 \citep{me02a,geb02}.
Targets were
acquired in imaging mode and placed into a 1$\farcs$2 $\times$
30$\arcsec$ slit.  Observations were made in sets of 5
exposures dithered 4--5$\arcsec$ along the slit, with individual
integrations of 300 sec per exposure.  B and A dwarf stars,
chosen for their lack of metal lines at moderate
resolution, were observed near the target objects for flux
calibration and telluric corrections. Spectral lamps reflected off
of the 4m dome spot were observed each night for pixel response
calibration, and a series of dark frames with identical exposure times
as the spectral flats were also obtained to remove
detector bias.

Data reduction consisted of initially trimming the
science images to eliminate vignetted
regions; dividing by a normalized, dark-subtracted flat field (constructed from
a median combination of the spectral lamp and bias exposures); and
correcting for bad pixels by interpolation,
using a mask created from flat-field and bias frames.
Images were then pairwise subtracted to eliminate sky
background and dark current.  Curvature of the dispersion lines was
determined by tracing the spectra of the standard stars, and this
trace was used as a template for the target spectra.
Both standard and object spectra
were extracted by summing 8--12 columns (depending on seeing conditions)
along each row after subtracting off the median background in that row.
Individual spectra from each order were then scaled by a multiplicative factor
and combined by averaging, rejecting 3$\sigma$ outliers in each
spectral bin. Wavelength calibration was done with the telluric OH lines, using
identifications from \citet{oli92}.  A telluric
correction was computed for each target/standard pair
by interpolating over H$_2$O absorption features in the standard star spectra and
ratioing these spectra with the uncorrected standard spectra.
A smoothed flux correction was then calculated by interpolating over Hydrogen
Paschen and Brackett lines
in the telluric-corrected standard spectra and
multiplying by the appropriate blackbody \citep{tok00}.  Finally,
the flux-calibrated spectral orders were combined by first scaling each
order to match overlap regions
(typically 1.28--1.30, 1.53--1.55, and 1.95--1.97 $\micron$), and
then correcting the combined spectra
to 2MASS H-band magnitudes.
Note that the overlap between the H- and
K-band orders falls within the 1.9 $\micron$ H$_2$O band; because of this,
we generally applied the same scaling corrections to both orders.

The majority of sources observed appear to be late-type (KM) background stars
based on the presence of weak H$_2$O absorption and relatively featureless
H-band spectra.
This is consistent with their positions on the near-infrared
color-color diagram in Figure 1 (which plots
revised photometry from
the 2MASS ADR\footnote{The ADR colors for many of our candidates are
outside of our original search constraints, which are based on
2MASS WPSD data.  These differences
are due to the improved photometric calibration of 2MASS J-band data in the ADR,
which compensates for short (hourly)
variations in telluric opacity that influence the broad J-band filter employed in this
survey.  See \citet{cut03} for further details.}), within the substantial
color uncertainties; c.f., the M dwarf and giant tracks of \citet{bes88}.
Furthermore, their faint magnitudes (15.5 ${\lesssim}$ $J$ ${\lesssim}$ 16)
imply optical/near-infrared
color limits ($R-J >$ 4--5, based on no detection in DSS photographic plates;
Reid et al.\ 1991) consistent with late-type stars.
In some cases these identifications were confirmed spectroscopically by comparison to
near-infrared data of known M dwarfs obtained in prior OSIRIS campaigns \citep{me02a}.
None of the brighter
background objects appear to be L dwarfs based on similar comparisons.  Adequate
spectral comparison for many of the spectra was not possible,
however, due to low signal-to-noise ratio data
($\sim$ 5-10 at J-band in the worst cases), although it is clear that these
faint objects are not T dwarfs.
We forego detailed classification
of background sources as it is beyond the scope of this paper,
and identify them simply as late-type stars.
We do, however, identify one candidate,
2MASS 2005$-$1056, as LHS 483, a DC9 white dwarf \citep[and references therein]{mco99}
with substantial proper motion
($\mu$ = 1$\farcs$08 yr$^{-1}$; Luyten 1979).  This object was not initially recognized as a background
proper motion star because of the small epoch difference between the available
DSS plates (2.1 yrs\footnote{We note that \citet{bak02} derive
$\mu = 6{\farcs}82$ yr$^{-1}$ for LHS 483 but cite the value as highly
uncertain because of the small DSS epoch difference.  From our imaging observations
and 2MASS data (see $\S$3.4) we derive $\mu = 0{\farcs}98{\pm}0{\farcs}03$ yr$^{-1}$,
roughly consistent with Luyten's measurement.}).  A second source, 2MASS 0533$-$0633, appears
to be a reddened, mid- to late-type M dwarf based
on its somewhat stronger H$_2$O bands, weak 1.23 $\micron$ FeH absorption, 1.25 $\micron$ K I doublet
lines, and red near-infrared spectral slope.  This source is also aligned with a small molecular
clump just south of the Orion Nebular Cloud \citep[ONC]{bal87,car01}.  Its late-type dwarf spectrum
and proximity to the 1--2 Myr ONC star-forming region \citep{hil97}
suggests that it could be a young brown dwarf.

\section{New T Dwarfs}

\subsection{Spectral Analysis}

We spectroscopically confirm three objects as T dwarfs: 2MASS 0348$-$6022,
2MASS 0516$-$0445, and
2MASS 2228$-$4310.  Astrometric and photometric properties are listed in Table 3, and
2MASS and DSS
images of each source (5$\arcmin$$\times$5$\arcmin$ field) are shown in Figure 2.
Calibrated spectral data
are shown in Figure 3, along with data for 2MASS 0243$-$2453, 2MASS 0415$-$0935,
and SDSS 0423$-$0414.  Note that data for 2MASS 0348$-$6022 and 2MASS 0516$-$0445
have been combined from multiple nights to improve signal-to-noise.
All six T dwarfs show distinct CH$_4$ bands at 1.1, 1.3 (blue and
red slopes of
the 1.27 $\micron$ J-band peak), 1.6, and 2.2 $\micron$; H$_2$O bands at 1.3 and
1.7 $\micron$; and K I absorption lines at 1.2432 and 1.2522 $\micron$.  The spectrum
of SDSS 0423$-$0414 shows some indication of CO absorption at 2.3 $\micron$.  The strength of the
CH$_4$ and H$_2$O bands in the three new discoveries are consistent with late-type
T dwarfs \citep{me02a,geb02}.

Figure 4 displays the 1.19 to 1.33 $\micron$ region of the spectra in
Figure 3, highlighting the 1.2432/1.2522 $\micron$ K I doublet lines.  These lines
are prominent in early- and mid-type T dwarf spectra (as well as late-type M
and L dwarfs), peaking in strength around spectral type T5--T5.5, then fading in
the later dwarfs \citep{me02a}.
We measured pseudo-equivalent widths\footnote{The presence of overlying opacity
throughout the near-infrared prevents the
measurement of ``true continuum'' for computing equivalent
widths; hence, the reported measurements are relative to the
local, or ``pseudo''-continuum.} (pEWs) for these lines by
integrating over the line profiles; see \citet{me02a}.  Results are listed in Table 4.
Values decrease monotonically from
roughly 12 to 4 {\AA} for the 1.2432
$\micron$ line and 16 to 6.5 {\AA} for the 1.2522 $\micron$ line
over spectral types T5.5 to T7, consistent with previously observed trends.
Upper pEW limits of roughly 2 {\AA}
are found for the K I lines of the latest-type T dwarf 2MASS 0415$-$0935.

\subsection{Classification}

We derived spectral types for the three discoveries using the
classification scheme of \citet{me02a}.  The top of
Table 5 lists spectral ratio values of observed T dwarf standards,
including new measurements for SDSS 0423$-$0414,
2MASS 0243$-$2453, and 2MASS 0415$-$0935 (our T0, T6, and T8 standards,
respectively), and measurements using
OSIRIS data for SDSS 1254$-$0122 \citep[T2]{leg00},
2MASS 0559$-$0414 \citep[T5]{me00c},
and 2MASS 0727+1710 \citep[T7]{me02a} from \citet{me02a}.  We make use of the
H$_2$O-B, CH$_4$-A, and CH$_4$-B indices only,
due to the wavelength limits of the spectral data
(obviating the the H$_2$O-A index), poor signal-to-noise at K-band
(obviating the CH$_4$-C, and 2.11/2.07 indices) and possible
uncertainties in the individual band calibration (obviating
the H/J and K/J color indices).  As can be seen in Table 5, the three remaining
indices are monotonic with spectral type
for the standards.

Spectral ratios of our discoveries are listed at the bottom of Table 5.
For each ratio, we assigned
a whole or half subtype based on the closest match to the standard values.
Final subtypes were derived from the average of these
individual ratio types, which for each object differ by less than $\pm$1 subclass.  Furthermore,
individual spectra for 2MASS 0348$-$6022 and 2MASS 0516$-$0445 obtained on separate nights
yield identical types within $\pm$0.5 subclasses.  Therefore, we assert that our classifications
are accurate to within 0.5 subclasses despite the limited suite of spectral ratios
employed.
Derived spectral types are T7 for 2MASS 0348$-$6022, T5.5 for 2MASS 0516$-$0445, and
T6.5 for 2MASS 2228$-$4310.  These mid- to late-type classifications are consistent with their
relatively blue near-infrared colors.

\subsection{Distance Estimates}

Distance estimates for our new discoveries can be made using their spectral
types, 2MASS photometry, and the absolute magnitudes of T dwarfs with measured parallaxes.  We first
compared 2MASS M$_J$, M$_H$, and M$_{K_s}$ versus spectral type (SpT) for T dwarfs typed T5 and later
using parallax data from \citet{dah02}; \citet{tin03}; and F.\ Vrba (2002, private communication).
Excluding the
known binaries 2MASS 1225$-$2739AB and 2MASS 1534$-$2952AB \citep{me03b} and Gliese 229B
(Nakajima et al.\ 1995; not detected by 2MASS),
we derive the following linear relations over the
range T5 to T8:
\begin{equation}
\begin{array}{rcl}
M_J & = & (10.00{\pm}0.12) + (0.84{\pm}0.02){\times}{\rm SpT}  \\
M_H & = & (9.60{\pm}0.18) + (0.88{\pm}0.03){\times}{\rm SpT}  \\
M_{K_s} & = & (8.7{\pm}0.3) + (0.98{\pm}0.05){\times}{\rm SpT}, \\
\end{array}
\end{equation}
where SpT(T5) = 5, SpT(T8) = 8, etc.
Based on these relations, we derived distance estimates for each source in each band assuming
spectral type uncertainties of $\pm$0.5 subclasses.  The mean distances and standard deviations are
9$\pm$4, 34$\pm$13, and 12$\pm$4 pc for 2MASS 0348$-$6022, 2MASS 0516$-$0445,
and 2MASS 2228$-$4310, respectively.
It is not surprising that the first object,
which has the latest classification and brightest apparent magnitudes, is likely to be
less than 10 pc from the Sun, unless it is an unresolved multiple system.
However, parallax observations are required to verify these
rather uncertain estimates.

\subsection{Proper Motions}

Proper motions for the T dwarf discoveries
were measured between the 2MASS and follow-up OSIRIS images following
the technique of \citet{me03c}.  First, astrometric
calibrations of each OSIRIS field, of the form
\begin{equation}
\begin{array}{rcl}
\alpha & = & {\alpha}_o + Ax + By \\
\delta & = & {\delta}_o + Cx + Dy, \\
\end{array}
\end{equation}
($x$ and $y$ are pixel coordinates on the OSIRIS images)
were made by linear regression using the 2MASS coordinates of
6--8 background sources
within 2$\farcm$5 of each T dwarf.  Background sources were verified to
show consistent positions between the two epochs (which span 3.1--4.1 yr)
to within three times the astrometric fit uncertainties,
roughly 0$\farcs$05--0$\farcs$1.   Positions for the T dwarfs
on the OSIRIS images were then calculated and compared to the original 2MASS coordinates.

Results are listed in Table 3.  As expected from their relative proximity, all of these
sources have statistically significant proper motions.
Their tangential velocities ($V_t$) are relatively modest, however,
ranging from 17 to 55 km s$^{-1}$.
These values are similar to the median for disk dwarfs (39 km s$^{-1}$;
Reid \& Hawley 2000) and
comparable to or less than the median for field late-type M and L dwarfs
(22 km s$^{-1}$; Gizis et al.\ 2000).  Hence, the kinematics of these three T dwarfs
are consistent with, but not restricted to, membership in the Galactic disk.

\section{Summary}

We have identified three new southern T dwarfs using the 2MASS survey.
These objects are important additions to the sample of nearby stars and
brown dwarfs; indeed, one object, if it is single, is likely to be closer than 10 pc from the Sun.
Spectroscopic observations yield classifications ranging from T5.5 to T7, and
measured K I pEWs support the observed trend of decreasing line strength with spectral
type for mid- to late-type T dwarfs.
All three objects have small to moderate proper motions and tangential
velocities, consistent with membership in the Galactic disk.
With roughly 70\% of the Southern Hemisphere portion of our 2MASS T Dwarf search
sample now completed and 13 T dwarfs so far identified \citep{me99,me00a,me00c,me02a},
we expect to uncover another 5--6 mid- to late-type T dwarfs over the remaining portion of
this part of the sky to $J = 16$.

\acknowledgments

We thank our Telescope Operator Sergio Pizarro
and Support Scientist Nicole van der Bliek for their assistance at the telescope,
and Kelle Cruz for her presence during the observations.
We also thank our anonymous referee for her/his rapid and thorough
examination of our submitted manuscript.
A.\ J.\ B.\ acknowledges support by NASA
through Hubble Fellowship grant HST-HF-01137.01 awarded by the
Space Telescope Science Institute, which is operated by the
Association of Universities for Research in Astronomy, Inc., for
NASA, under contract NAS 5-26555. This research has made use of
the SIMBAD database, operated at CDS, Strasbourg, France.
AAO and SERC
images were obtained from the Digitized Sky Survey
image server maintained by the Canadian Astronomy Data Centre,
which is operated by the Herzberg Institute of Astrophysics,
National Research Council of Canada.
This publication makes use of data from the Two
Micron All Sky Survey, which is a joint project of the University
of Massachusetts and the Infrared Processing and Analysis Center,
funded by the National Aeronautics and Space Administration and
the National Science Foundation.
2MASS data were obtained through
the NASA/IPAC Infrared Science Archive, which is operated by the
Jet Propulsion Laboratory, California Institute of Technology,
under contract with the National Aeronautics and Space
Administration.

\clearpage

\begin{deluxetable}{llccccl}
\tabletypesize{\scriptsize}
\tablecaption{T Dwarf Candidates Absent in Follow-up OSIRIS Images.}
\tablewidth{0pt}
\tablehead{
 & & \multicolumn{4}{c}{2MASS Observations\tablenotemark{b}} & \\
\cline{3-6}
\colhead{Object\tablenotemark{a}} &
\colhead{$\beta$ ($\degr$)} &
\colhead{UT Date} &
\colhead{$J$} &
\colhead{$J-H$} &
\colhead{$H-K_s$} &
\colhead{Identification\tablenotemark{c}} \\
\colhead{(1)} &
\colhead{(2)} &
\colhead{(3)} &
\colhead{(4)} &
\colhead{(5)} &
\colhead{(6)} &
\colhead{(7)} \\}
\startdata
2MASS J01315876$-$1259435 & $-$21 & 2000 Oct 23 & 15.86$\pm$0.07 & 0.14$\pm$0.16 & 0.15$\pm$0.24 &   \\
2MASS J01392545$-$0258486 & $-$12 & 1998 Sep 23 & 15.68$\pm$0.06 & 0.46$\pm$0.12 & $-$0.20$\pm$0.20 & 2001 KY8 \\
2MASS J01432550$-$1240101 & $-$22 & 2000 Oct 10 & 15.79$\pm$0.07 & 0.40$\pm$0.11 & $-$0.14$\pm$0.21 &  \\
2MASS J01492365$-$1945149 & $-$29 & 1998 Aug 10 & 15.53$\pm$0.05 & 0.29$\pm$0.09 & 0.15$\pm$0.14 &  \\
2MASS J01542969$-$1258435 & $-$23 & 1998 Aug 20 & 15.50$\pm$0.05 & 0.53$\pm$0.08 & $-$0.13$\pm$0.16 & 2000 EY156 \\
2MASS J02061785+0456321 & $-$7 & 2000 Nov 27 & 15.96$\pm$0.08 & 0.06$\pm$0.21 & 0.21$\pm$0.28 & 2000 SX124 \\
2MASS J02100930+0908316 & $-$4 & 2000 Sep 28 & 15.76$\pm$0.06 & 0.37$\pm$0.11 & $-$0.04$\pm$0.22 &  \\
2MASS J02253536$-$5539594 & $-$63 & 1999 Oct 27 & 15.30$\pm$0.08 & $-$0.20$\pm$0.16 & $<$ 0.86 & Image artifact \\
2MASS J02525674$-$8020549 & $-$72 & 2000 Oct 04 & 15.83$\pm$0.10 & $-$0.03$\pm$0.21 & $<$ 0.94 & Image artifact \\
2MASS J03014529$-$6055462 & $-$70 & 1999 Nov 11 & 15.33$\pm$0.10 & $-$0.44$\pm$0.21 & $<$ 1.37 & Image artifact \\
2MASS J03184819$-$4216275 & $-$57 & 1999 Sep 26 & 15.14$\pm$0.09 & $-$0.90$\pm$0.24 & $<$ 0.86 & Image artifact \\
2MASS J03295950$-$5446545 & $-$69 & 1999 Nov 02 & 15.38$\pm$0.11 & 0.17$\pm$0.22 & $<$ 1.34 & Image artifact \\
2MASS J03451647+1132394 & $-$8 & 2000 Nov 25 & 15.77$\pm$0.07 & 0.57$\pm$0.11 & $-$0.13$\pm$0.17 & 1990 QB3 \\
2MASS J03594765+1341458 & $-$7 & 1998 Nov 17 & 15.51$\pm$0.06 & 0.47$\pm$0.09 & $-$0.15$\pm$0.16 &  \\
2MASS J19084533$-$3352087 & $-$11 & 2000 Jul 20 & 15.86$\pm$0.06 & 0.26$\pm$0.13 & 0.31$\pm$0.19 &  \\
2MASS J19183027$-$3427382 & $-$12 & 1998 Jul 25 & 15.63$\pm$0.05 & 0.03$\pm$0.14 & 0.39$\pm$0.20 &  \\
2MASS J19231778$-$4832099 & $-$26 & 2000 Oct 12 & 15.71$\pm$0.09 & $-$0.16$\pm$0.22 & $<$ 0.60 & Image artifact \\
2MASS J19243027$-$4332438 & $-$21 & 1999 Jul 14 & 15.96$\pm$0.09 & 0.26$\pm$0.18 & 0.35$\pm$0.22 & 1996 TP34 \\
2MASS J19263511$-$3435565 & $-$13 & 1998 Jul 26 & 15.89$\pm$0.09 & 0.05$\pm$0.21 & 0.38$\pm$0.29 &  \\
2MASS J19355972$-$2923351 & $-$8 & 1998 Jul 05 & 15.91$\pm$0.09 & 0.22$\pm$0.17 & $-$0.01$\pm$0.28 &  \\
2MASS J19492452$-$3201418 & $-$11 & 2000 Aug 10 & 15.66$\pm$0.06 & 0.53$\pm$0.11 & $-$0.03$\pm$0.17 & 2001 RZ \\
2MASS J19530053$-$3512193 & $-$14 & 1999 Jul 15 & 15.52$\pm$0.06 & 0.47$\pm$0.09 & $-$0.26$\pm$0.15 &  \\
2MASS J20153100$-$2958197 & $-$10 & 1998 Aug 10 & 15.09$\pm$0.03 & 0.29$\pm$0.07 & 0.17$\pm$0.12 &  \\
2MASS J20180855$-$3250195 & $-$13 & 1999 Sep 26 & 15.80$\pm$0.07 & 0.21$\pm$0.14 & $-$0.15$\pm$0.28 &  \\
2MASS J20203953$-$3414209 & $-$14 & 1999 Jun 13 & 15.89$\pm$0.09 & 0.22$\pm$0.16 & 0.14$\pm$0.25 &  \\
2MASS J20342244$-$3748127 & $-$18 & 1998 Sep 08 & 15.83$\pm$0.06 & 0.40$\pm$0.11 & $-$0.11$\pm$0.21 & 1989 TR2 \\
2MASS J20352362$-$3858012 & $-$20 & 1998 Sep 08 & 15.80$\pm$0.05 & 0.45$\pm$0.10 & $-$0.22$\pm$0.21 &  \\
2MASS J20464656$-$2831162 & $-$10 & 2000 Jul 24 & 15.38$\pm$0.05 & 0.29$\pm$0.09 & 0.03$\pm$0.14 &  \\
2MASS J20493283$-$4532443 & $-$27 & 1999 Jul 26 & 15.97$\pm$0.07 & 0.19$\pm$0.15 & 0.37$\pm$0.22 &  \\
2MASS J20495554$-$3525083 & $-$17 & 1999 Sep 09 & 15.96$\pm$0.09 & 0.27$\pm$0.15 & 0.24$\pm$0.21 &  \\
2MASS J20514464$-$3123264 & $-$13 & 1999 Jul 09 & 15.93$\pm$0.06 & 0.38$\pm$0.11 & $-$0.26$\pm$0.27 &  \\
2MASS J20533768$-$3227270 & $-$14 & 1998 Aug 20 & 15.60$\pm$0.06 & 0.62$\pm$0.09 & $-$0.14$\pm$0.14 & 1999 XN168 \\
2MASS J20572107$-$3304186 & $-$15 & 1998 Aug 20 & 16.00$\pm$0.07 & 0.56$\pm$0.14 & $-$0.23$\pm$0.23 &  \\
2MASS J21035085$-$2237412 & $-$6 & 2000 Aug 06 & 15.46$\pm$0.05 & 0.25$\pm$0.10 & 0.05$\pm$0.16 &  \\
2MASS J21050753$-$3122506 & $-$14 & 1998 Aug 20 & 15.71$\pm$0.05 & 0.39$\pm$0.10 & $-$0.02$\pm$0.18 &  \\
2MASS J21054515$-$3542535 & $-$18 & 1998 Aug 20 & 15.67$\pm$0.04 & 0.58$\pm$0.08 & $-$0.01$\pm$0.13 &  \\
2MASS J21090201$-$3725168 & $-$20 & 1999 Aug 12 & 15.53$\pm$0.08 & 0.18$\pm$0.14 & 0.38$\pm$0.18 &  \\
2MASS J21113838$-$4133320 & $-$24 & 1999 Sep 10 & 15.74$\pm$0.06 & 0.37$\pm$0.11 & $-$0.06$\pm$0.19 &  \\
2MASS J21165349$-$3325598 & $-$17 & 1998 Aug 15 & 15.80$\pm$0.06 & 0.43$\pm$0.14 & $-$0.09$\pm$0.22 & 1999 WA1 \\
2MASS J21170863$-$2441328 & $-$8 & 1998 Aug 04 & 15.38$\pm$0.05 & 0.49$\pm$0.07 & $-$0.02$\pm$0.12 & 2001 FP26 \\
2MASS J21183794$-$2813093 & $-$12 & 1998 Aug 04 & 15.85$\pm$0.06 & 0.57$\pm$0.11 & $-$0.07$\pm$0.19 &  \\
2MASS J21285884$-$4249292 & $-$26 & 1999 Aug 20 & 15.46$\pm$0.06 & 0.39$\pm$0.08 & $-$0.02$\pm$0.14 &  \\
2MASS J21413332$-$2453233 & $-$10 & 1998 Aug 12 & 15.73$\pm$0.06 & 0.39$\pm$0.09 & $-$0.16$\pm$0.16 &  \\
2MASS J22050171$-$2658019 & $-$14 & 2000 Jul 24 & 15.41$\pm$0.05 & 0.27$\pm$0.08 & $-$0.19$\pm$0.16 &  \\
2MASS J22154587$-$2605584 & $-$14 & 2000 Jul 24 & 15.53$\pm$0.05 & 0.29$\pm$0.09 & 0.04$\pm$0.15 &  \\
2MASS J22165310$-$2656329 & $-$15 & 1998 Jul 31 & 15.43$\pm$0.06 & 0.05$\pm$0.13 & 0.09$\pm$0.20 &  \\
2MASS J22311396$-$3417156 & $-$23 & 2000 Jul 24 & 14.53$\pm$0.03 & 0.29$\pm$0.05 & 0.00$\pm$0.07 &  \\
2MASS J22385056$-$5456370 & $-$42 & 2000 Aug 02 & 15.98$\pm$0.09 & 0.13$\pm$0.19 & 0.21$\pm$0.27 &  \\
2MASS J22420219$-$3346098 & $-$24 & 1999 Jul 26 & 15.56$\pm$0.05 & 0.43$\pm$0.09 & $-$0.10$\pm$0.17 & 1997 YZ3 \\
2MASS J22541137$-$4826135 & $-$38 & 1999 Sep 21 & 15.97$\pm$0.08 & 0.14$\pm$0.13 & 0.35$\pm$0.23 &  \\
\enddata
\tablenotetext{a}{All objects are listed with their 2MASS ADR Point Source Catalog source
designations, given as ``2MASS Jhhmmss[.]ss$\pm$ddmmss[.]s''. The
suffix conforms to IAU nomenclature convention and is the
sexagesimal Right Ascension and declination at J2000 equinox.}
\tablenotetext{b}{Photometry from the 2MASS ADR; note that some objects have revised photometry
placing them outside of our original WPSD search constraints.}
\tablenotetext{c}{Asteroid identifications are from the Small-Body
Search Tool maintained by the Jet Propulsion Laboratory Solar System Dynamics Group:
\url{http://ssd.jpl.nasa.gov/cgi-bin/sb{\_}search}.}
\end{deluxetable}

\begin{deluxetable}{lccccclcl}
\rotate
\tabletypesize{\scriptsize}
\tablecaption{Log of OSIRIS Spectroscopic Observations.}
\tablewidth{0pt}
\tablehead{
\colhead{Object} &
\colhead{$J$\tablenotemark{a}} &
\colhead{$J-K_s$\tablenotemark{a}} &
\colhead{UT Date} &
\colhead{t (s)} &
\colhead{Airmass} &
\colhead{Calibrator} &
\colhead{SpT} &
\colhead{Identification} \\
\colhead{(1)} &
\colhead{(2)} &
\colhead{(3)} &
\colhead{(4)} &
\colhead{(5)} &
\colhead{(6)} &
\colhead{(7)} &
\colhead{(8)} &
\colhead{(9)} \\}
\startdata
2MASS J00231718$-$3346366 & 16.01$\pm$0.08 & 0.61$\pm$0.23 & 2002 Sep 23 & 1500 & 1.00 & HD 6619 & A1 V & Late-type star \\
2MASS J02053685$-$0853442 & 15.89$\pm$0.06 & 0.73$\pm$0.17 & 2002 Sep 20 & 1500 & 1.09 & HD 13936 & A0 V & Late-type star \\
2MASS J02431371$-$2453298 & 15.38$\pm$0.05 & 0.16$\pm$0.17 & 2002 Sep 24 & 1500 & 1.04 & HD 20293 & A5 V & T6 standard \\
2MASS J02485356$-$4951573 & 16.14$\pm$0.12 & 1.13$\pm$0.19 & 2002 Sep 22 & 1500 & 1.14 & HD 17098 & B9 V & Reddened star \\
2MASS J02512393$-$3842100 & 15.95$\pm$0.08 & 0.78$\pm$0.19 & 2002 Sep 22 & 1500 & 1.03 & HD 17098 & B9 V & Late-type star \\
2MASS J02560785$-$4311108 & 15.69$\pm$0.06 & 0.75$\pm$0.17 & 2002 Sep 24 & 1500 & 1.04 & HD 20293 & A5 V & Late-type star \\
2MASS J03164047$-$6202104 & 15.85$\pm$0.08 & 1.02$\pm$0.14 & 2002 Sep 22 & 1500 & 1.19 & HD 22252 & B8 V & Late-type star \\
2MASS J03480772$-$6022270 & 15.32$\pm$0.06 & $-$0.28$\pm$0.24 & 2002 Sep 21 & 1500 & 1.19 & HD 24863 & A4 V & T dwarf \\
2MASS J04143647$-$6916594 & 15.72$\pm$0.07 & 0.89$\pm$0.16 & 2002 Sep 24 & 1500 & 1.31 & HD 37935 & B9.5 V & Late-type star \\
2MASS J04151954$-$0935066 & 15.70$\pm$0.06 & 0.27$\pm$0.21 & 2002 Sep 24 & 1500 & 1.08 & HD 28763 & A2/3 V & T8 standard \\
SDSSp J042348.57$-$041403.5 & 14.47$\pm$0.03 & 1.54$\pm$0.04 & 2002 Sep 23 & 1500 & 1.29 & HD 30321 & A2 V & T0 standard \\
2MASS J05160945$-$0445499 & 15.98$\pm$0.09 & 0.50$\pm$0.22 & 2002 Sep 21 & 1500 & 1.12 & HD 33948 & B5 V & T dwarf \\
 & & & 2002 Sep 22 & 1500 & 1.11  & HD 37303  & B1 V  &  \\
 & & & 2002 Sep 24 & 1500 & 1.11  & HD 28763  & A2/3 V  &  \\
2MASS J05335973$-$0633098 & 15.76$\pm$0.06 & 0.93$\pm$0.10 & 2002 Sep 22 & 1500 & 1.16 & HD 37303 & B8 V & Reddened M dwarf\tablenotemark{b} \\
2MASS J17065753$-$0428539 & 15.96$\pm$0.09 & 0.53$\pm$0.20 & 2002 Sep 23 & 1500 & 1.38 & HD 159170 & A5 V & Late-type star \\
2MASS J18031154+1305281 & 16.03$\pm$0.09 & 0.85$\pm$0.18 & 2002 Sep 22 & 3000 & 1.52 & HD 171975 & B9 V & Late-type star \\
2MASS J19092267$-$8234478 & 15.52$\pm$0.06 & 0.74$\pm$0.13 & 2002 Sep 22 & 1500 & 1.67 & HD 169904 & B8 V & Late-type star \\
2MASS J19215777$-$6303224 & 15.62$\pm$0.05 & 0.66$\pm$0.14 & 2002 Sep 20 & 3000 & 1.21 & HD 186837 & B5 V & Late-type star \\
2MASS J19300878$-$4435012 & 15.96$\pm$0.08 & 1.06$\pm$0.14 & 2002 Sep 21 & 1500 & 1.03 & HD 176425 & A0 V & Late-type star \\
2MASS J19322282$-$6835592 & 16.00$\pm$0.09 & 0.97$\pm$0.16 & 2002 Sep 24 & 1500 & 1.28 & HD 184586 & A1 V & Late-type star \\
2MASS J19325619$-$4851162 & 15.99$\pm$0.09 & 1.01$\pm$0.15 & 2002 Sep 23 & 1500 & 1.25 & HD 189388 & A2/3 V & Late-type star \\
2MASS J19383909$-$2735379 & 15.97$\pm$0.08 & 0.62$\pm$0.21 & 2002 Sep 21 & 1500 & 1.02 & HD 176425 & A0 V & Late-type star \\
2MASS J19394892$-$5531025 & 15.87$\pm$0.08 & 0.51$\pm$0.18 & 2002 Sep 20 & 1500 & 1.19 & HD 186837 & B5 V & Late-type star \\
2MASS J19465571$-$3644491 & 15.58$\pm$0.07 & 0.72$\pm$0.14 & 2002 Sep 21 & 1500 & 1.07 & HD 176425 & A0 V & Late-type star\tablenotemark{c} \\
2MASS J20014023$-$4111011 & 15.94$\pm$0.06 & 0.51$\pm$0.20 & 2002 Sep 20 & 1500 & 1.33 & HD 189399 & A2/3 V & Late-type star \\
2MASS J20053482$-$1056545 & 15.28$\pm$0.05 & 0.53$\pm$0.12 & 2002 Sep 22 & 1500 & 1.20 & HD 202753 & B5 V & LHS 483, white dwarf \\
2MASS J20364476+0335475 & 15.75$\pm$0.08 & 0.46$\pm$0.19 & 2002 Sep 23 & 1500 & 1.20 & HD 198070 & A0 Vn & Late-type star \\
2MASS J20425201$-$7924433 & 15.67$\pm$0.06 & 0.64$\pm$0.15 & 2002 Sep 21 & 1500 & 1.77 & HD 203955 & A0 V & Late-type star \\
2MASS J20592033+1752232 & 16.02$\pm$0.09 & 0.79$\pm$0.20 & 2002 Sep 24 & 1500 & 1.60 & HD 207563 & B2 V & Late-type star \\
2MASS J21264358$-$0926573 & 15.91$\pm$0.09 & 0.86$\pm$0.17 & 2002 Sep 22 & 1500 & 1.24 & HD 215143 & A0 Vn & Late-type star \\
2MASS J21301473$-$0720578 & 15.48$\pm$0.08 & 0.66$\pm$0.16 & 2002 Sep 20 & 1500 & 1.14 & HD 186837 & B5 V & Late-type star \\
2MASS J21393009$-$0928268 & 15.88$\pm$0.07 & 0.53$\pm$0.20 & 2002 Sep 22 & 1500 & 1.31 & HD 215143 & A0 Vn & Late-type star \\
2MASS J21443131+0327100 & 15.84$\pm$0.06 & 1.02$\pm$0.13 & 2002 Sep 23 & 1500 & 1.21 & HD 198070 & A0 Vn & Late-type star \\
2MASS J22282889$-$4310262 & 15.66$\pm$0.08 & 0.37$\pm$0.22 & 2002 Sep 24 & 3000 & 1.10 & HD 220802 & B9 V & T dwarf \\
\enddata
\tablenotetext{a}{Photometry from the 2MASS ADR; note that some objects have revised photometry
outside of our original WPSD search constraints.}
\tablenotetext{b}{Candidate young brown dwarf in the Orion A complex \citep{str93,car01}.}
\tablenotetext{c}{This object appeared much fainter at J-band when imaged with OSIRIS than
as observed by 2MASS.  It is possible that the 2MASS source is a chance alignment with an uncatalogued
minor planet, or that this source is intrinsically variable.}
\end{deluxetable}

\begin{deluxetable}{llccccccc}
\tabletypesize{\footnotesize}
\tablecaption{Observed Properties of the New T Dwarfs.}
\tablewidth{0pt}
\tablehead{
\colhead{Object} &
\colhead{Type} &
\colhead{$J$} &
\colhead{$J-H$} &
\colhead{$H-K_s$} &
\colhead{$d_{sp}$\tablenotemark{a}} &
\colhead{$\mu$} &
\colhead{$\theta$} &
\colhead{$V_t$} \\
\colhead{} &
\colhead{} &
\colhead{} &
\colhead{} &
\colhead{} &
\colhead{(pc)} &
\colhead{($\arcsec$ yr$^{-1}$)} &
\colhead{($\degr$)} &
\colhead{(km s$^{-1}$)} \\
\colhead{(1)} &
\colhead{(2)} &
\colhead{(3)} &
\colhead{(4)} &
\colhead{(5)} &
\colhead{(6)} &
\colhead{(7)} &
\colhead{(8)} &
\colhead{(9)} \\}
\startdata
2MASS J03480772$-$6022270 & T7 & 15.32$\pm$0.06 & $-$0.24$\pm$0.16 & $-$0.04$\pm$0.28 & 9$\pm$4 & 0.77$\pm$0.04 & 201$\pm$3 &  32$\pm$14 \\
2MASS J05160945$-$0445499 & T5.5 & 15.98$\pm$0.09 & 0.26$\pm$0.19 & 0.24$\pm$0.27 &  34$\pm$13 &  0.34$\pm$0.03 & 232$\pm$5 & 55$\pm$21 \\
2MASS J22282889$-$4310262 & T6.5 & 15.66$\pm$0.08 & 0.30$\pm$0.14 & 0.07$\pm$0.24 & 12$\pm$4 &  0.31$\pm$0.03 & 175$\pm$15 & 17$\pm$6 \\
\enddata
\tablenotetext{a}{Estimated spectrophotometric distance based on spectral type and
2MASS $JHK_s$ photometry; see $\S$ 3.3.}
\end{deluxetable}

\begin{deluxetable}{llccccc}
\tabletypesize{\small}
\tablecaption{K I Pseudo-equivalent Widths.}
\tablewidth{0pt}
\tablehead{
 & &
\multicolumn{2}{c}{1.2432 $\micron$} & &
\multicolumn{2}{c}{1.2522 $\micron$} \\
\cline{3-4} \cline{6-7}
\colhead{Object} &
\colhead{Type} &
\colhead{$\lambda$$_c$ ($\micron$)} &
\colhead{pEW ({\AA})} & &
\colhead{$\lambda$$_c$ ($\micron$)} &
\colhead{pEW ({\AA})} \\
\colhead{(1)} &
\colhead{(2)} &
\colhead{(3)} &
\colhead{(4)} &  &
\colhead{(5)} &
\colhead{(6)} \\}
\startdata
SDSS 0423$-$0414 & T0 & 1.243 & 6.4$\pm$0.8 & & 1.252 & 6.6$\pm$1.0 \\
2MASS 0516$-$0445\tablenotemark{a} & T5.5 & 1.242 & 11.6$\pm$2.3 & & 1.252 & 15.9$\pm$2.0  \\
2MASS 0243$-$2453 & T6 & 1.243 & 6.8$\pm$1.8 & & 1.252 & 10.2$\pm$1.1  \\
2MASS 2228$-$4310 & T6.5 & 1.243 & 5.0$\pm$1.0 & & 1.251 & 9.9$\pm$1.6  \\
2MASS 0348$-$6022 & T7 & 1.242 & 4.0$\pm$1.1 & & 1.250 & 6.4$\pm$1.2  \\
2MASS 0415$-$0935 & T8 & --- & $<$ 1.7 & & --- & $<$ 2.2 \\
\enddata
\tablenotetext{a}{Measured from combined spectrum.}
\end{deluxetable}

\begin{deluxetable}{lcccl}
\tabletypesize{\small}
\tablecaption{Spectral Ratios and Classification on the \citet{me02a} Scheme.}
\tablewidth{0pt}
\tablehead{
\colhead{Object} &
\colhead{H$_2$O-B} &
\colhead{CH$_4$-A} &
\colhead{CH$_4$-B} &
\colhead{SpT\tablenotemark{a}} \\
\colhead{(1)} &
\colhead{(2)} &
\colhead{(3)} &
\colhead{(4)} &
\colhead{(5)} \\}
\startdata
\cline{1-5}
\multicolumn{5}{c}{Standards} \\
\cline{1-5}
SDSS 0423$-$0414 & 0.689 & 0.985 & 0.916 &  T0 \\
SDSS 1254$-$0122\tablenotemark{b} & 0.559 & 0.943 & 0.826 &  T2 \\
2MASS 0559$-$1404\tablenotemark{b} & 0.456 & 0.789 & 0.383  &  T5  \\
2MASS 0243$-$2453 & 0.369 & 0.648 & 0.229 &  T6  \\
2MASS 0727$-$1710\tablenotemark{b} & 0.332 & 0.509 & 0.133 &  T7  \\
2MASS 0415$-$0935 & 0.267 & 0.429 & 0.062 &  T8  \\
\cline{1-5}
\multicolumn{5}{c}{Discoveries} \\
\cline{1-5}
2MASS 0348$-$6022 & 0.371(6) & 0.541(7) & 0.104(7/8) &  T7 \\
2MASS 0516$-$0445\tablenotemark{c} & 0.378(6) & 0.812(5) & 0.207(6) &   T5.5  \\
2MASS 2228$-$4310 & 0.285(7/8) & 0.681(6) & 0.228(6) &  T6.5 \\
\enddata
\tablenotetext{a}{Adopted spectral type for standards; derived spectral type for discoveries,
with uncertainty $\pm$0.5 subclasses.}
\tablenotetext{b}{Data from \citet{me02a}.}
\tablenotetext{c}{Measured from combined spectrum.}
\end{deluxetable}

\clearpage

\begin{figure}
\epsscale{0.8}
\plotone{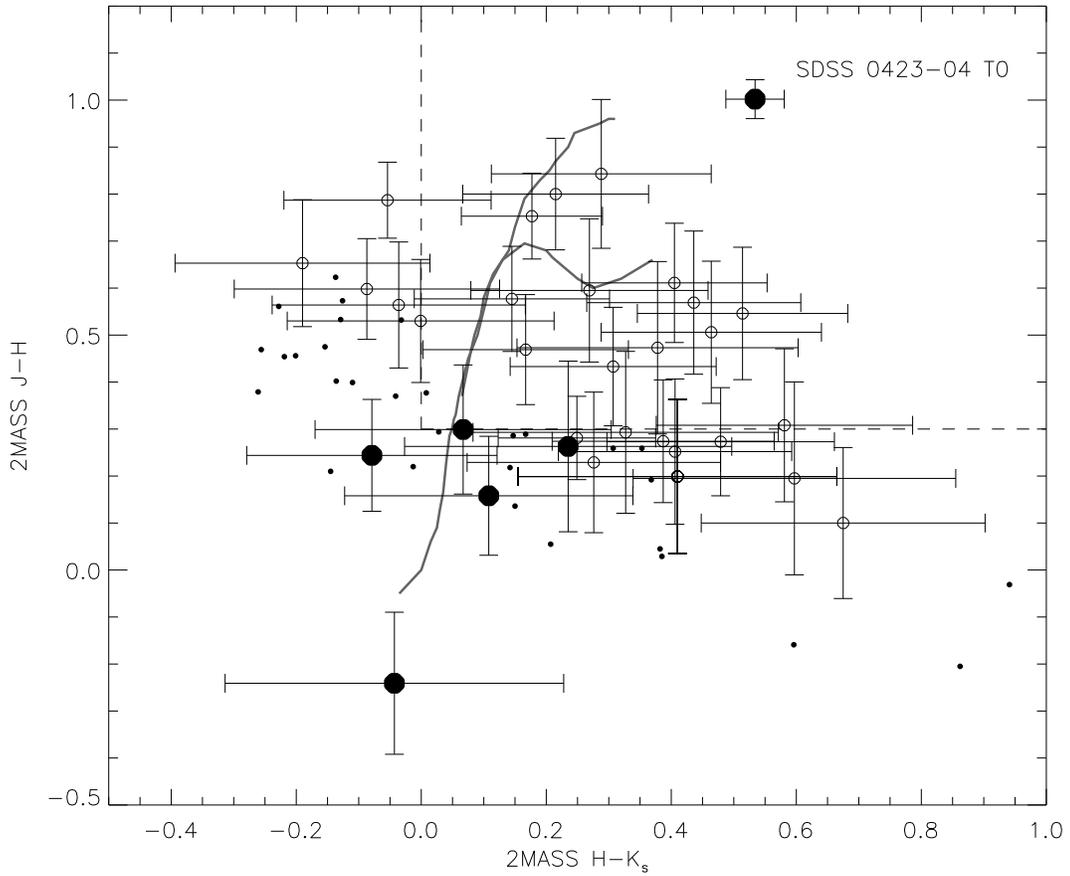}
\caption{2MASS near-infrared color-color diagram for observed T dwarfs and T dwarf candidates,
with photometry from the 2MASS ADR.
Small points indicate objects absent in follow-up
OSIRIS images; open circles indicate background sources based on spectroscopic
observations; large filled circles indicate the three new T dwarfs and
the known T dwarfs 2MASS 0243$-$2453, 2MASS 0415$-$0935, and SDSS 0423$-$0414
(identified).
Dwarf (lower) and giant (upper) tracks from \citet{bes88} are indicated by grey lines.  Note
that many of the background sources have revised 2MASS ADR
colors outside of our original WPSD search constraints
(dashed lines).  The revised colors are
consistent with their spectroscopic identifications as late-type background stars.
 \label{fig0} }
\end{figure}

\clearpage

\begin{figure}
\epsscale{0.8}
\caption{Findercharts for the T dwarf discoveries, showing AAO/SERC (R-band),
and 2MASS ($J$- and $K_s$-bands) fields.  Images are scaled to the same
spatial resolution, 5$\arcmin$ on a side, with North up and East
to the left.  A 10$\arcsec$ box is centered on the position of
the T dwarf in all images.  $Note: Figure 2 is included as a separate jpeg file due
to space considerations.$ \label{fig1} }
\end{figure}

\clearpage

\begin{figure}
\plotone{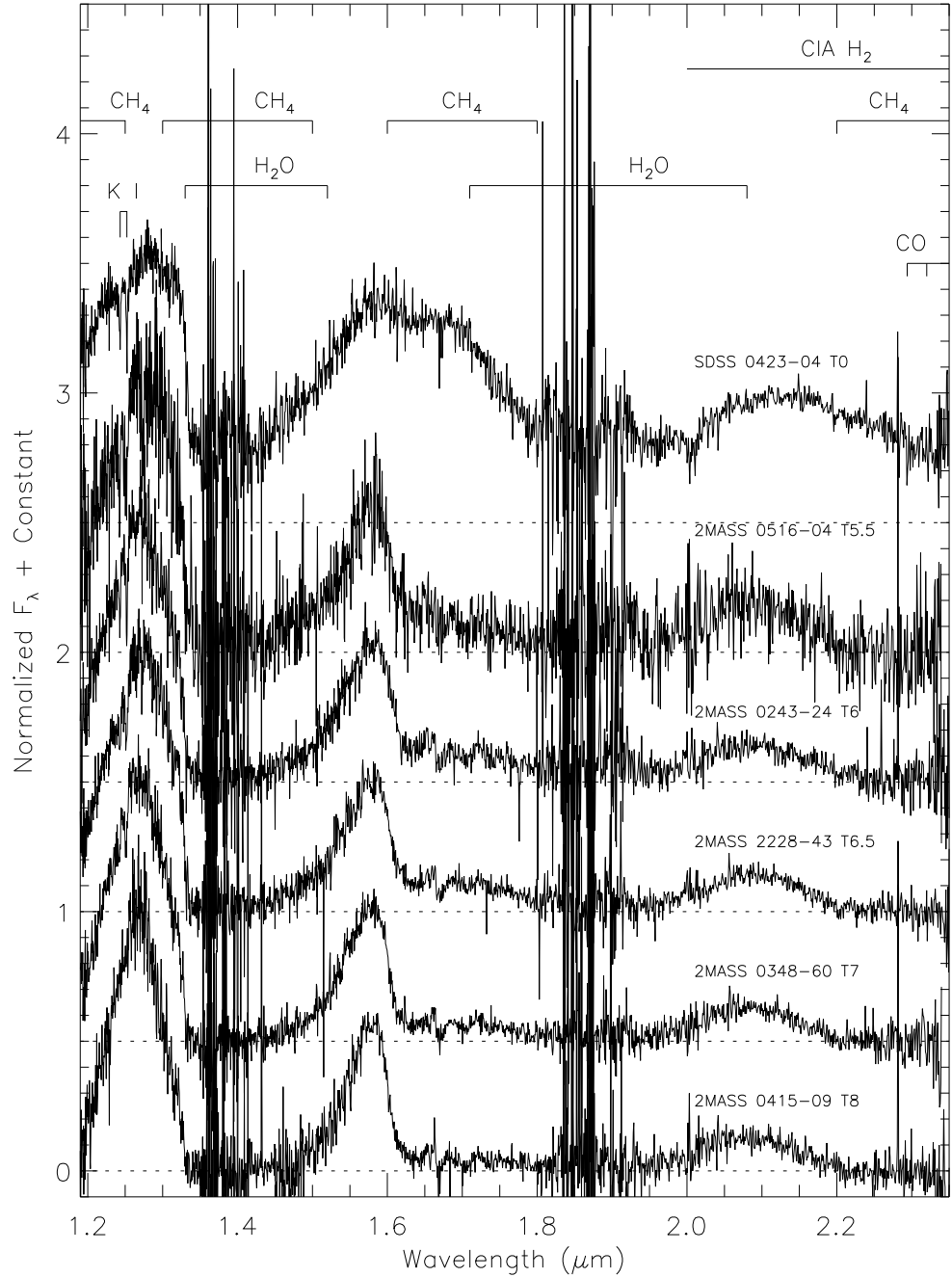}
\caption{Reduced OSIRIS spectra of the observed T dwarfs in order of spectral type.  Data
are normalized at 1.2 $\micron$ and offset by a constant (dashed line).
Major features of CH$_4$, H$_2$O, CO, and K I are identified, as well as the
region most strongly affected by CIA H$_2$ absorption.  \label{fig2} }
\end{figure}

\clearpage

\begin{figure}
\epsscale{0.5}
\plotone{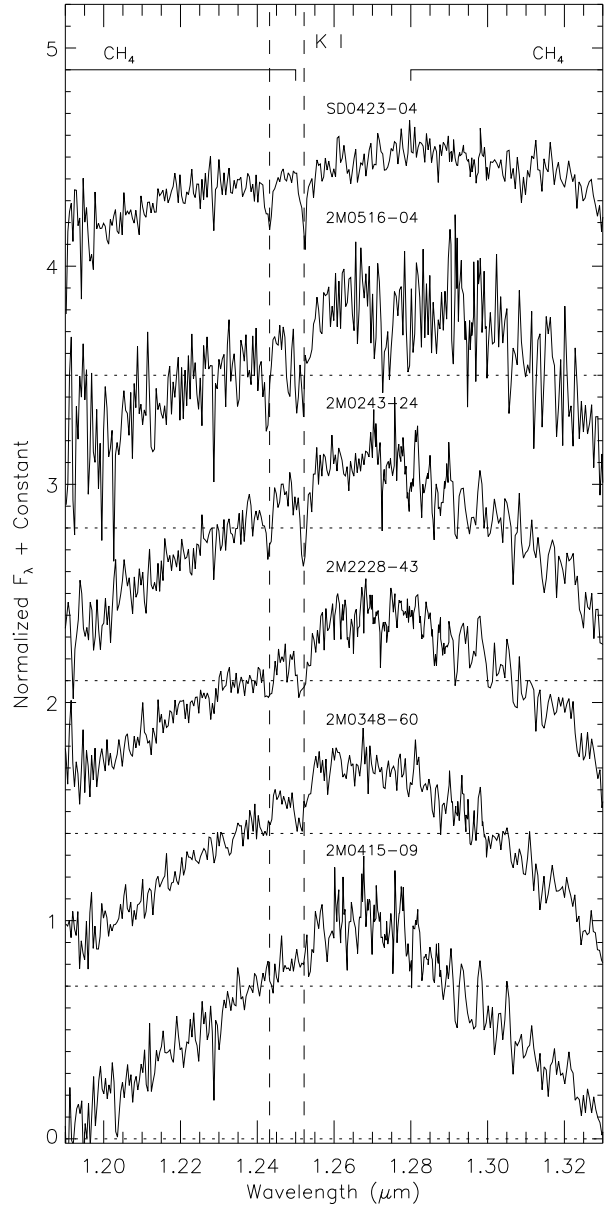}
\caption{OSIRIS spectral data for the observed T dwarfs from 1.19 to 1.33 $\micron$,
highlighting the 1.2432/1.2522 $\micron$ K I doublet (vertical dashed lines).  Spectra are
scaled and offset as in Figure 3.  Note that the feature at 1.27 $\micron$ in
the spectra of 2MASS 0516$-$0445, 2MASS 0243$-$2453, and 2MASS 2228$-$4310
is an artifact of the data reduction.
\label{fig3}}
\end{figure}

\end{document}